\documentclass[12pt]{article}
\title{Hamiltonian Lattice QCD with Wilson Fermions at Strong Coupling}
\author{
Yi-Zhong Fang
and Xiang-Qian Luo\thanks{Corresponding author. E-mail: stslxq@zsu.edu.cn}
\vspace{2cm}
\\
{\small\sl Department of Physics, Zhongshan University,
Guangzhou 510275, People's Repulic of China
}\\
}

\begin{document}
\maketitle

\begin{flushleft}
{\bf Abstract}
\end{flushleft}
\noindent
Hamiltonian lattice QCD with Wilson fermions is investigated systematically
by strong-coupling
expansion up to the second order. The effective Hamiltonian is diagonalized
by Bogoliubov transformation. The vacuum energy, chiral condensate,
pseudo-scalar and vector meson masses are calculated. The comparison with
the unitary transformation method by Luo and Chen is also made. The method
discussed in this paper has potential application to QCD at finite density.

\newpage

\section{Introduction}
Lattice Gauge Theory (LGT) proposed by Wilson \cite{wilson}
is the most promising
non-perturbative method for QCD. In comparison to other techniques,
the advantage of LGT is that there is no free parameter
when the continuum limit is taken. There are two different
approaches:
Lagrangian formulation and Hamiltonian formulation.
Lagrangian formulation is convenient for Monte Carlo simulations.
Hamiltonian formulation is useful for analytical calculations
\cite{Kogut75,smit,yaouance,luo,luo2,yasuo}.
Although the standard lattice Lagrangian Monte Carlo method
has been successful in many aspects of LGT,
it unfortunately breaks down at finite chemical
potential due to the ``complex
action problem''. On the other hand, lattice QCD at finite chemical
potential
formulated in the Hamiltonian approach does not encounter the complex action
problem.
Recently, Gregory, Guo, Kr\"oger and Luo \cite{gregory} investigated
the finite density Hamiltonian lattice QCD with naive fermions,
using the unitary transformation, variational method
and Bogoliubov transformation method developed by Luo and Chen \cite{luo}.

  It is well known that naive fermions experience ``species doubling''.
Kogut-Susskind fermions and Wilson fermions are the most popular
ways towards solving of the problem.
Although the doubling problem has not yet been completely solved,
Kogut-Susskind fermions are frequently used for
exploring the spontaneous breaking of the U(1) chiral symmetry.
For Wilson fermions,  the doubling modes are removed,
but chiral symmetry
is explicitly broken and fine tuning of the fermion mass remains to be done;
This method is very popular in spectrum computations, due to the existence
of
the flavor symmetry.

  In this paper, we study the Hamiltonian lattice QCD with Wilson fermions
by
strong coupling expansion method. The effective Hamiltonian is
Fierz-rearranged,
re-expressed by meson operators, and then diagonalized
by Bogoliubov transformation.
The vacuum energy, chiral condensate, pseudo-scalar and vector meson masses
are computed.

The rest of the paper is organized as follows. In Sect. \ref{our approach},
we obtain the effective
Hamiltonian using strong coupling expansion.
The physical results are presented
in Sect. \ref{results} and Sect. \ref{result2}.
The conclusion is provided in Sect. \ref{discussion}.

\section{Effective Hamiltonian at strong coupling}
\label{our approach}

We begin with the (d+1)-dimensional lattice QCD Hamiltonian
with Wilson fermions
\begin{eqnarray}
H &=& \left(m+{rd \over a}\right)\sum_{x} \bar{\psi}(x)\psi(x)
+{1 \over 2a}\sum_{x}\sum_{k=\pm1}^{\pm
d}\bar{\psi}(x)\gamma_{k}U(x,k)\psi(x+k)
\nonumber \\
&-& {r \over 2a} \sum_{x}\sum_{k=\pm1}^{\pm d}\bar{\psi}(x)U(x,k)\psi(x+k)
+ {g^{2} \over 2a}
\sum_{y}\sum_{j=1}^{d} E^{\alpha}_{j}(x)E^{\alpha}_{j}(x)\nonumber \\
&-& {1 \over ag^{2}} \sum_{p} {\rm Tr} \left(U_{p}+U_{p}^{+}-2\right),
\label{first}
\end{eqnarray}
\\
where $m$, $a$, $r$ and $g$ are respectively the bare fermion mass, lattice
spacing,
Wilson parameter, and bare coupling constant. $U(x,k)$ is the gauge link
variable at site $x$ and direction $k$,
and $\psi$ is the four-component
spinor. The convention $\gamma_{-k}=-\gamma_{k}$ is used.
$E^{\alpha}(x)$ is the color-electric field and
summation over $\alpha=1,2, ..., 8$ is implied.
$U_p$ is the product of gauge link variable around an elementary plaquette
and in the
continuum, it represents the color magnetic interactions.
However, at the strong coupling,
the color magnetic energy (the last term)  can be ignored.

We want to diagonalize $H$ so that the fermion field $\psi$ can be expressed
in terms of up and down 2-spinors $\xi$ and $\eta^{\dagger}$
\begin{eqnarray}
\psi(x)=\left(
\begin{array}{c}
\xi(x)\\
\eta^{+}(x)
\end{array}
\right).
\end{eqnarray}
The bare vacuum state $\vert 0 \rangle$ is defined by
$\xi|0\rangle=\eta|0\rangle=E_{j}^{\alpha}(x)|0\rangle=0$.
Since the up and down components are coupled via the $\gamma_{k}$ matrices,
the bare vacuum is not an eigenstate of $H$.

For the convenience of strong coupling expansion, we introduce the
un-perturbative
term $W_0$ and perturbative term $W$, i.e.,
\begin{eqnarray}
{2a \over g^2} H = W_0+W,
\end{eqnarray}
where
\begin{eqnarray}
 W_0 &=& (m+\frac {rd}{a})\frac {2a}{g^2} \sum_{x}\bar{\psi}(x)\psi(x)
+\sum_{x,j}E^{\alpha}_{j}(x)E^{\alpha}_{j}(x),
\nonumber \\
W &=& \frac {1}{g^2} \sum_{x,k}\bar{\psi}(x)(-r+\gamma_{k})U(x,k)\psi(x+k).
\label{second}
\end{eqnarray}
The ground state energy is $g^2 \varepsilon/2a$, where for $1/g^2 <<1$,
$\varepsilon$
can be approximated by
\begin{eqnarray}
\varepsilon = \langle 0|\frac {2a}{g^2}H|0\rangle
=  \varepsilon^{(0)}+\varepsilon^{(1)}+\varepsilon^{(2)}.
\label{third}
\end{eqnarray}
$\varepsilon^{(0)}$, $\varepsilon^{(1)}$ and $\varepsilon^{(2)}$
are the zeroth, first and second order corrections:
\begin{eqnarray}
\varepsilon^{(0)} &=& W_{0}|0\rangle,
\nonumber \\
\varepsilon^{(1)} &=& \langle 0|W|0\rangle,
\nonumber \\
\varepsilon^{(2)} &=& \langle 0|W\frac {1-|0\rangle \langle
0|}{\varepsilon^{(0)}
-W_{0}} W|0\rangle.
\label{forth}
\end{eqnarray}
From (\ref{second}) and (\ref{forth}), we obtain
\begin{eqnarray}
\varepsilon^{(1)} &=& 0,
\nonumber \\
\varepsilon^{(0)}
&=&
\frac {2aM}{g^2} \langle 0|\sum_{x}{\bar \psi}(x)\psi (x) |0\rangle
=-2N_c N_fN_s\frac {2aM}{g^2}.
\label{fifth}
\end{eqnarray}
Here $N_s$, $N_{c}$
and $N_{f}$ are respectively
the number of lattice sites,  colors and  flavors.
$M$
\begin{equation}
M=m+ {rd \over a},
\label{fifth_1}
\end{equation}
and $1/(2Ma)$ is an analog of the $\kappa$ parameter in Lagrangian LGT.
Using
\begin{eqnarray}
W_{0}W|0\rangle &=& [W_{0}, W]|0\rangle+WW_{0}|0\rangle
\nonumber\\
&=& [W_{0}, W]|0\rangle+\varepsilon^{(0)}W|0\rangle,
\label{seventh}
\end{eqnarray}
the second order correction becomes
\begin{equation}
\varepsilon^{(2)}=\langle 0|W\frac {1-|0\rangle \langle
0|}{\varepsilon^{(0)}
-W_0}W|0\rangle
=-{1 \over C_N}\langle 0|W \left(1-|0\rangle \langle 0|\right)W|0\rangle
=-{1 \over C_N} \langle 0| WW |0\rangle,
\label{eighth}
\end{equation}
where $C_{N}={\rm Tr} \sum_{\alpha}\lambda^{\alpha}
\lambda^{\alpha}/N_C=(N_{C}^{2}-1)/(2N_{C})$,
i.e.  the
Casimir invariant. From (\ref{fifth}) and (\ref{eighth}), we have
\begin{eqnarray}
\varepsilon =\varepsilon^{(0)}-\langle 0|\frac {WW}{C_{N}}|0\rangle =\langle
0 \vert {2a \over g^2} H_{eff} \vert 0 \rangle.
\end{eqnarray}
Therefore, the effective Hamiltonian is
\begin{eqnarray}
H_{eff} = M\sum_{x}\bar{\psi}(x)\psi(x)+H_{r}^{eff}+H_{k}^{eff}.
\label{ninth}
\end{eqnarray}
This equation will serve as our starting point, where
\begin{equation}
H_{r}^{eff}=-\frac {Kr^2}{2aN_{C}}\sum_{x,k}\bar{\psi}_{c_{1},f_{1}}
(x)\psi_{c_{2},f_{1}}(x+k)\bar{\psi}_{c_{2},f_{2}}(x+k)\psi_{c_{1},f_{2}}(x)
,
\label{eleventh}
\end{equation}
and
\begin{equation}
H_{k}^{eff}=\frac {K}{2aN_{C}}\sum_{x,k}\bar{\psi}_{c_{1},f_{1}}
(x)\gamma_{k}\psi_{c_{2},f_{1}}(x+k)\bar{\psi}_{c_{2},f_{2}}(x+k)\gamma_{k}
\psi_{c_{1},f_{2}}(x).
\label{twelth}
\end{equation}
$K=1/(g^{2}C_{N})$ is the effective coupling of the four-fermion
interactions.
The color index $c$ and flavor index $f$ are
explicitly written, and summation over repeated index is implied.
It is easily proven that the contribution from terms proportional
to $O(r)$ vanishes.

\begin{table}
\begin{center}
\begin{tabular}{|c|c|c|c|c|c|c|c|c|}
\hline
$\Gamma_A$ & 1 & $i\gamma_j$ & $\gamma_4$ & $i\gamma_5$
& $\gamma_4 \gamma_5$ & $\gamma_4 \gamma_j$ &
$i \epsilon_{j j_1 j_2} \gamma_{j_1} \gamma_{j_2}$ &
$i \epsilon_{j j_1 j_2} \gamma_4 \gamma_{j_1} \gamma_{j_2}$\\
\hline
$L_A'$ & 1 & 1 & 1 & 1 & 1 & 1 &1 &1\\
\hline
\end{tabular}
\end{center}
\caption{$\Gamma$ matrices and coefficients for $H_r^{eff}$.}
\end{table}

\begin{table}
\begin{center}
\begin{tabular}{|c|c|c|c|c|c|c|c|c|}
\hline
$\Gamma_A$ & 1 & $i\gamma_j$ & $\gamma_4$ & $i\gamma_5$
& $\gamma_4 \gamma_5$ & $\gamma_4 \gamma_j$ &
$i \epsilon_{j j_1 j_2} \gamma_{j_1} \gamma_{j_2}$ &
$i \epsilon_{j j_1 j_2} \gamma_4 \gamma_{j_1} \gamma_{j_2}$\\
\hline
$L_A$ & -1 & $1-2\delta_{k,j}$ & 1 & 1 & -1 & $-1+2\delta_{k,j}$ &
$1-2\delta_{k,j}$ &
$-1+2\delta_{k,j}$\\
\hline
\end{tabular}
\end{center}
\caption{$\Gamma$ matrices and coefficients for $H_k^{eff}$.}
\end{table}

In order to express the effective Hamiltonian by meson operators,
we perform a Fierz transformation so that (\ref{eleventh}) and
(\ref{twelth})
are rearranged as,
\begin{equation}
H_{r}^{eff}=-\frac {Kr^{2}d}{a}\sum_{x}\psi^{\dagger}(x)\psi (x)+
\frac {Kr^{2}}{8aN_{c}}\sum_{x,k}L_{A}^{'}\psi_{f_{1}}^{\dagger}(x)
\Gamma_{A}\psi_{f_{2}}(x)\psi_{f_{2}}^{\dagger}(x+k)\Gamma_{A}\psi_{f_{1}}(x
+k),
\label{thirteenth}
\end{equation}
and
\begin{equation}
H_{k}^{eff}=-\frac {Kd}{a}\sum_{x}\psi^{\dagger}(x)\psi (x)-\frac
{K}{8aN_{c}}\sum_{x,k}L_{A}\psi^{\dagger}_{f_{1}}(x)\Gamma_{A}
\psi_{f_{2}}(x)\psi^{\dagger}_{f_{2}}(x+k)\Gamma_{A}\psi_{f_{1}}(x+k).
\label{forteenth}
\end{equation}
where the matrices $\Gamma_A$ and
their coefficients $L'_A$ and $L_A$ are given in Tab. [1] and Tab.[2],
and summation over the index $A$ is understood.
Note our basis of the sixteen $\Gamma_A$ matrices is the same as
the standard reference \cite{peskin}, but it is different from
Smit\cite{smit}.

Substituting (\ref{thirteenth}) and
(\ref{forteenth}) into (\ref{ninth}), we have
\begin{eqnarray}
H_{eff} &=& M\sum_{x}\bar{\psi}(x)\psi (x)-\frac {K}{a}(r^{2}+1)d
\sum_{x}\psi^{\dagger}(x)\psi (x)
\nonumber \\
&+&\frac {K}{8aN_c}\sum_{x}
\sum_{k=\pm j}
\bigg[ (r^{2}+1)
\psi^{\dagger}(x)\psi (x)\psi^{\dagger}(x+k)\psi (x+k)
\nonumber \\
&+& (r^{2}-1)\psi^{\dagger} (x)\gamma_{4}\psi (x)
\psi^{\dagger}(x+k)\gamma_{4}\psi (x+k)
\nonumber \\
&- & (r^{2}-1)\psi^{\dagger}(x)\gamma_{5}\psi
(x)\psi^{\dagger}(x+k)\gamma_{5}
\psi (x+k)
\nonumber \\
&+&
(r^{2}+1)\psi^{\dagger}(x)\gamma_{4}\gamma_{5}\psi (x)
 \psi^{\dagger}(x+k)\gamma_{4}\gamma_{5}\psi (x+k)
\nonumber \\
& +&
\left(r^{2}+(1-2\delta_{k,j})\right)
\psi^{\dagger}(x)\gamma_{4}\gamma_{j}\psi (x)
\psi^{\dagger}(x+k)\gamma_{4}\gamma_{j}\psi (x+k)
\nonumber \\
&- &
\left( r^{2}-(1-2\delta_{k,j})\right)\psi^{\dagger}(x)\gamma_{j}\psi
(x)\psi^{\dagger}(x+k)\gamma_{j}\psi (x+k)
\nonumber \\
&- &
\left((1-2\delta_{k,j})+r^{2}\right)
\psi^{\dagger}(x)\gamma_{4}\sigma_{j}\psi (x)
\psi^{\dagger}(x+k)\gamma_{4}\sigma_{j}\psi (x+k)
\nonumber \\
&+& \left((1-2\delta_{k,j})
- r^{2}\right)\psi^{\dagger}(x)\sigma_{j}\psi (x)\psi^{\dagger}(x+k)
\sigma_{j}\psi (x+k)\bigg],
\label{fifteenth}
\end{eqnarray}
where $\sigma_{j}=\epsilon_{jj_{1}j_{2}}\gamma_{j_{1}}\gamma_{j_{2}}$.
This effective Hamiltonian describes the nearest-neighbor four fermion
interactions. For
$r \to 0$, $H_{eff}$ is the same as that from the unitary transformation
method using by Luo and Chen\cite{luo}.

Define the pseudo-scalar meson operator $\Pi$ and the vector meson operator
$V$ as \cite{gregory}
\begin{eqnarray}
\Pi(x)_{f_{1}f_{2}} &=& \frac {1}{2\sqrt{2N_{c}}}\psi^{\dagger}_{f_{1}}(x)
(1-\gamma_{4})\gamma_{5}\psi_{f_{2}}(x),
\nonumber\\
\Pi^{\dagger}(x)_{f{2}f{1}} &=& \frac
{1}{2\sqrt{2N_{c}}}\psi^{\dagger}_{f_{2}}
(x)(1+\gamma_{4})\gamma_{5}\psi_{f_{1}}(x),
\nonumber\\
V_{j}(x)_{f_{1}f_{2}} &=& \frac
{1}{2\sqrt{2N_{c}}}\psi^{\dagger}_{f_{1}}(x)(1-
\gamma_{4})\gamma_{j}\psi_{f_{2}}(x),
\nonumber\\
V^{\dagger}_{j}(x)_{f_{2}f_{1}} &=& \frac {1}{2\sqrt{2N_{c}}}\psi^{\dagger}_
{f_{2}}(x)(1+\gamma_{4})\gamma_{j}\psi_{f_{1}}(x).
\label{sixteenth}
\end{eqnarray}
Then Eq. (\ref{fifteenth}) can be re-expressed as
\begin{eqnarray}
H_{eff} &=& (-2N_{c}N_{f}V)\left( M+\frac {Kd}{a} \right)
+ 2M\sum_{x,f_{1}f_{2}j}
\left( \Pi^{\dagger}_{f_{2}f_{1}}(x)\Pi_{f_{1}f_{2}}(x)
+V^{\dagger}_{jf_{2}f_{1}}(x)V_{jf_{1}f_{2}}(x)\right)
\nonumber\\
&-&\frac {2Kr^{2}d}{a}\sum_{xf_{1}f_{2}j}\left(\Pi^{\dagger}_{f_{2}f_{1}}
(x)\Pi_{f_{1}f_{2}}(x)
+V^{\dagger}_{jf_{2}f_{1}(x)} V_{jf_{1}f_{2}}(x)\right)
\nonumber\\
&-& \frac {Kr^{2}}{2a}\sum_{xf_{1}f_{2}kj} \bigg(
\Pi^{\dagger}_{f_{1}f_{2}}(x)\Pi_{f_{2}f_{1}}(x+k)
+\Pi_{f_{1}f_{2}}(x)\Pi^{\dagger}_{f_{2}f_{1}}(x+k)
\nonumber\\
&+& V^{\dagger}_{jf_{1}f_{2}}(x)V_{jf_{2}f_{1}}(x+k)
+V_{jf_{1}f_{2}}(x)V^{\dagger}_{jf_{2}f_{1}}(x+k) \bigg)
\nonumber\\
&+& \frac {K}{2a}\sum_{x,k}
\bigg(\Pi^{\dagger}_{f_{1}f_{2}}(x)\Pi^{\dagger}_{f_{2}f_{1}}(x+k)
+\Pi_{f_{1}f_{2}}(x)\Pi_{f_{2}f_{1}}(x+k)
\nonumber \\
&+& \left( V^{\dagger}_{jf_{1}f_{2}}(x)V^{\dagger}_{jf_{2}f_{1}}(x+k)
+V_{jf_{1}f_{2}}(x)V_{jf_{2}f_{1}}(x+k)\right) (1-2\delta_{k,j})\bigg)
\nonumber\\
&+ & \frac
{2Kd}{a}\left(\sum_{xf_{1}f_{2}}\Pi^{\dagger}_{f_{2}f_{1}}(x)\Pi_{f_{1}f_{2}
}(x)+\sum_{xf_{1}f_{2}j}V^{\dagger}_{jf_{2}f_{1}}(x)V_{jf_{1}f_{2}}(x)
\right)
\nonumber \\
&=& (-2N_{c}N_{f}N_s) \left(M+\frac {Kd}{a} \right)+H_{\Pi}+H_{V},
\label{seventeenth}
\end{eqnarray}
where the contributions of the pseudo-scalar mesons are vector mesons are
respectively
\begin{eqnarray}
H_{\Pi} &=& \left(-\frac {2Kr^{2}d}{a}+2M \right)
\sum_{xf_{1}f_{2}}\Pi^{\dagger}_{f_{2}f_{1}}(x)\Pi_{f_{1}f_{2}}(x)
\nonumber \\
&-& \frac {Kr^{2}}{2a}\sum_{xf_{1}f_{2}k}
\left(\Pi^{\dagger}_{f_{1}f_{2}}(x)\Pi_{f_{2}f_{1}}(x+k)
+\Pi_{f_{1}f_{2}}(x)\Pi^{\dagger}_{f_{2}f_{1}}(x+k)\right)
\nonumber \\
&+& \frac {K}{2a}\sum_{xf_{1}f_{2}}\left(\Pi^{\dagger}_{f_{1}f_{2}}(x)
\Pi^{\dagger}_{f_{2}f_{1}}(x+k)+\Pi_{f_{1}f_{2}}(x)\Pi_{f_{2}f_{1}}(x+k)
\right)
\nonumber \\
&+& \frac {2Kd}{a}\sum_{xf_{1}f_{2}}\Pi^{\dagger}_{f_{2}f_{1}}(x)
\Pi_{f_{1}f_{2}}(x),
\label{twentyfirst}
\end{eqnarray}
and
\begin{eqnarray}
H_{V} &=& \left(2M-\frac {2Kr^{2}d}{a}+\frac {2Kd}{a}\right)
\sum_{xf_{1}f_{2}j}
V^{\dagger}_{jf_{2}f_{1}}(x)V_{jf_{1}f_{2}}(x)
\nonumber \\
&-& \frac
{Kr^{2}}{2a}\sum_{xf_{1}f_{2}kj}\left(V^{\dagger}_{jf_{1}f_{2}}(x)V_{jf_{2}f
_{1}}(x+k)
+V_{jf_{1}f_{2}}(x)V^{\dagger}_{jf_{2}f_{1}}(x+k)\right)
\nonumber \\
&+& \frac {K}{2a}\sum_{xf_{1}f_{2}kj}\left(V^{\dagger}_{jf_{1}f_{2}}(x)
V^{\dagger}_{jf_{2}f_{1}}(x+k)+V_{jf_{1}f_{2}}(x)V_{jf_{2}f_{1}}(x+k)\right)
(1-2\delta_{k,j}).
\label{twentysecond}
\end{eqnarray}

\section{Meson masses}
\label{results}

After a Fourier transformation
\begin{equation}
\Pi_{f_{1}f_{2}}(x)=\sum_{p}e^{ipx}a_{f_{1}f_{2}}(p),
\label{twentythird}
\end{equation}
$H_{\Pi}$ in (\ref{twentyfirst}) becomes
\begin{eqnarray}
H_{\Pi} &=& \left(\frac {2Kd}{a}(1-r^{2})+2M\right)
\sum_{pf_{1}f_{2}} a^{\dagger}_{f_{1}f_{2}} (p)a_{f_{2}f_{1}}(p)
\nonumber \\
&-& \frac {Kr^{2}}{a}\sum_{f_{1}f_{2}}\sum_{p}
\left(
a^{\dagger}_{f_{1}f_{2}} (p)a_{f_{2}f_{1}}(p)
+a(p)_{f_{1}f_{2}}a^{\dagger}_{f_{2}f_{1}}(p) \right)
\sum_{j} \cos{p_{j}a}
\nonumber \\
&+&
\frac {K}{a}\sum_{pf_{1}f_{2}}
\left(
a^{\dagger}_{f_{1}f_{2}} (p)a^{\dagger}_{f_{2}f_{1}} (-p)
+a_{f_{1}f_{2}} (-p) a_{f_{2}f_{1}}(p)
\right)
\sum_{j} \cos{p_{j}a}.
\label{twentyforth}
\end{eqnarray}
The Bogoliubov transformation
\begin{eqnarray}
a(p) & \rightarrow & a(p)\cosh{u(p)}+a^{\dagger} (-p)\sinh{u(p)},
\nonumber \\
a^{\dagger} (p) & \rightarrow & a^{\dagger}(p)\cosh{u(p)}+a(-p)\sinh{u(p)},
\nonumber \\
a(-p) & \rightarrow & a(-p)\cosh{u(p)} + a^{\dagger}(p)\sinh{u(p)},
\nonumber \\
a^{\dagger}(-p) & \rightarrow & a^{\dagger}(-p) \cosh{u(p)} + a(p)
\sinh{u(p)},
\label{twentyfifth}
\end{eqnarray}
will lead to the diagonalization of  $H_{\Pi}$, if
the parameter $u_p$ satisfies
\begin{equation}
\tanh{2u(p)}=-\frac {2G_{2}}{G_{1}}\sum_{q=1}^d \cos{p_{q}a},
\label{twentysixeth}
\end{equation}
where
\begin{eqnarray}
G_{2} &=& \frac {K}{a},
\nonumber \\
 G_{1} &=& 2\left[M+\frac {Kd}{a}(1-r^{2})-\frac {Kr^{2}}{a}
\sum_{q=1}^d\cos{p_{q}a}\right].
\label{twentyseventh}
\end{eqnarray}
Therefore, after the Bogoliubov transformation, the diagonalized $H_{\Pi}$
becomes
\begin{eqnarray}
H_{\Pi} &=&
G_{1} \sum_{pf_{1}f_{2}} \left(1-\tanh^{2}{2u(p)}\right)^{\frac {1}{2}}
a^{\dagger}_{f_{1}f_{2}} (p) a_{f_{2}f_{1}} (p)
\nonumber \\
&-& \frac
{G_{1}}{2}N^{2}_{f}\sum_{p}\left[\left(1-\left(1-\tanh^{2}{2u(p)}\right)
^{\frac {1}{2}}\right)
+ \frac {2G_{2}r^{2}}{G_{1}}\sum_{q=1}^d \cos{p_{q}a}\right].
\label{twentyeighth}
\end{eqnarray}
From (\ref{twentysixeth}), (\ref{twentyseventh}) and (\ref{twentyeighth}),
we obtain the dispersion law for the pseudo-scalar mesons,
i.e. the relation between the energy $E_{\Pi}$ of the pseudo-scalar meson
and momentum $p$:
\begin{eqnarray}
E_{\Pi} &=& G_{1} \left(1-\tanh^{2}{2u(p)}\right)^{\frac {1}{2}}
= G_{1}\left[1-(\frac {2G_{2}d}{G_{1}})^{2}\right]^{\frac {1}{2}}
\nonumber \\
&=& 2\left[M+\frac {Kd}{a}(1-r^{2})-\frac {K(r^{2}-1)}{a} \sum_{q=1}^d
\cos{p_{q}a}\right]^{\frac {1}{2}}
\nonumber \\
& \times & \left[M+\frac {Kd}{a}(1-r^{2})
-\frac {K}{a}(r^{2}+1)\sum_{q=1}^d\cos{p_{q}a}\right]^{\frac {1}{2}}.
\label{twentyninth}
\end{eqnarray}
For naive fermions ($r=0$) in 3+1 dimensions,
according to  (\ref{twentyninth}),
we reproduce the result of Luo and Chen\cite{luo}:
$E^{2}_{\Pi}\vert_{p_{q}=0}=m_{\Pi}^2 \propto m$, i.e. the PCAC
theorem; In the chiral limit $m \to 0$,
the dispersion law tells us
\begin{equation}
E_{\Pi}^{2}=\frac {4K^{2}}{a^{2}}\left[9-(\sum_{q=1}^d
\cos{p_{q}a})^{2}\right].
\label{thirtieth}
\end{equation}
One sees that there exist two zeroes at $\vec{p}=(0,0,0)$ and
$\vec{p}=(\pi,\pi,\pi)$. This
is a well known ``doubling problem''.
For Wilson fermions ($r\not=0$), the doubler modes are removed,
but the chiral symmetry is explicitly broken.
In order to define the chiral limit, one
has to fine-tune $M \to M_c$ so that the pseudo-scalars become massless.
From (\ref{twentyninth}), we get
\begin{equation}
M_{c}=\frac {6Kr^{2}}{a}.
\label{thirtyfirst}
\end{equation}
According to this formulae and (\ref{twentyninth}), near the chiral limit,
the mass of a pseudo-scalar  behaves as
\begin{eqnarray}
 m_{\Pi}^2  &=& E^{2}_{\Pi}\vert_{p_q=0}
=4(M-M_{c})^{2}+\frac {24K}{a}(M-M_{c})
\nonumber \\
&\approx^{M\to M_{c}}&
\frac {24K}{a}(M-M_{c}),
\label{thirtythird}
\end{eqnarray}
which is the PCAC relation for a pseudo-scalar in the Wilson fermion case.

We now consider $H_{V}$ in (\ref{twentysecond})
similarly. After a Fourier transformation
\begin{equation}
V_{j}(x)=\sum_{p}e^{ipx}b_j (p),
\label{thirtyforth}
\end{equation}
and a Bogoliubov transformation
\begin{eqnarray}
b_j (p) & \rightarrow &  b_j (p)\cosh{v_j(p)}
+b^{\dagger}_j(-p)\sinh{v_j(p)},
\nonumber \\
b^{\dagger}_j (p) & \rightarrow & b^{\dagger}_j (p) \cosh{v_j(p)} +
b_j (-p) \sinh{v_j(p)},
\label{thirtyfifth}
\end{eqnarray}
where
\begin{equation}
\tanh{2v_j (p)}=-\frac {2G_{2}}{G_{1}}\left(\sum_{q=1}^d \cos{p_{q}a}-
2\cos{p_j}a \right),
\label{thirtysixeth}
\end{equation}
we obtain
\begin{eqnarray}
H_{V} &=& \sum_{pjf_{1}f_{2}}
G_{1}\left(1-\tanh^{2}{2v_j (p)} \right)^{\frac
{1}{2}}b^{\dagger}_{jf_{1}f_{2}} (p) b_{jf_{2}f_{1}} (p)
\nonumber \\
&-& \frac {G_{1}}{2}N^{2}_{f}\sum_{pj}
\left[\left(1-\left(1-\tanh^{2}{2v_j (p)}\right)^\frac {1}{2}\right)
+\frac {2G_{2}}{G_{1}}r^{2}\sum_{q=1}^d \cos{p_{q}a}\right].
\label{thirtyseventh}
\end{eqnarray}
Therefore, using (\ref{thirtyfirst}), the dispersion relation for the vector
mesons is
\begin{eqnarray}
E_{j} &=& \frac {2K}{a}\left[3(1+r^{2})+(1-r^{2})\sum_{q=1}^d\cos{p_{q}a}-
2\cos{p_{j}a}\right]^{\frac {1}{2}}
\nonumber \\
& \times & \left[3(1+r^{2})
-(1+r^{2})\sum_{q=1}^d \cos{p_{q}a}+2\cos{p_{j}a}\right]^{\frac {1}{2}}.
\label{thirtyeighth}
\end{eqnarray}
For naive fermions ($r=0$) in 3+1 dimensions,
from (\ref{thirtyfirst}) and (\ref{thirtyeighth}),
\begin{equation}
E^{2}_{j}=\frac {4K^{2}}{a^{2}}
\left[9-\left(\sum_{q=1}^d\cos{p_{q}a}-2\cos{p_{j}a}\right)\right],
\label{fortysecond}
\end{equation}
where the chiral limit has been taken.
One sees that there exist six zeroes on the boundary of  the first Brillouin
zone,
$\vec{p}=(\pi,0,0), (0,\pi,0), (0,0,\pi), (\pi,\pi,0), (0,\pi,\pi),
(\pi,0,\pi)$,
due to the ``doubling problem''.
For Wilson fermions ($r\not=0$), the doubling problem is avoided,
and the mass for a vector meson is
\begin{eqnarray}
m_{V}&=&E_{j}\vert_{p=0}
=G_{1}\left(1-\tanh^{2} 2v_j(p)\right)^{\frac {1}{2}} \vert_{p=0}
\nonumber \\
&=&G_{1}\left[1
-\left(\frac {2G_{2}}{G_{1}}(d-2)\right)^{2}\right]^{\frac
{1}{2}}\vert_{p=0},
\label{fortieth}
\end{eqnarray}
where (\ref{thirtysixeth}) has been used.
In the chiral limit $M = M_{c}$,
\begin{equation}
m_{V}=\frac {4\sqrt{2}}{a}K,
\label{fortyfirst}
\end{equation}
which is independent of the Wilson parameter $r$.

\section{Vacuum energy and chiral condensate}
\label{result2}

The vacuum energy is the vacuum expectation value of the Hamiltonian
\begin{eqnarray}
E_{\Omega} &=& \langle\Omega|H|\Omega\rangle=\langle 0|H_{eff}|0\rangle
\nonumber \\
 &=& (-2N_{c}N_{f}N_s)\left(M+\frac {d}{g^{2}C_{N}a}\right)
\nonumber \\
&-& \frac {G_{1}}{2} N^{2}_{f}
\sum_{p}\left[ \left(1-\left(1-\tanh^{2}{2u(p)}\right)^{\frac {1}{2}}\right)
+ \frac {2G_{2}r^{2}}{G_{1}}\sum_{q=1}^d\cos{p_{q}a}\right]
\nonumber \\
&-& \frac {G_{1}}{2}N^{2}_{f}\sum_{pj}
\left[\left(1-\left(1-\tanh^{2}{2v_j (p)}\right)^\frac {1}{2}\right)
+\frac {2G_{2}}{G_{1}}r^{2}\sum_{q=1}^d \cos{p_q a}\right].
\label{fortythird}
\end{eqnarray}
In the $r=0$ limit, this agrees with Luo and Chen\cite{luo}.

The fermion condensate
\begin{equation}
\langle \bar{\psi}\psi \rangle=-2N_{c}N_{f}N_s
+\langle \bar{\psi}\psi \rangle_{\Pi}+\langle \bar{\psi}\psi \rangle_{V}
\label{fortytorth}
\end{equation}
can also be computed using the Feynman-Hellmann theorem.
Here
\begin{eqnarray}
\langle \bar{\psi}\psi\rangle_{\Pi} &=& -\frac {N^{2}_{f}}{2}
\sum_{p}
\left[
{\frac {\partial G_{1}}{\partial m}}-\frac {1}{2}\frac {{\frac
{\partial}{\partial
m}}\left(G^{2}_{1}-(2G_{2}\sum_{q=1}^d\cos{p_{q}a})^{2}\right)}
{\sqrt{G^2_{1}-(2G_{2}\sum_{q=1}^d\cos{p_{q}a})^{2}}}
+{\frac {\partial G_{2}}{\partial m}}2r^{2}\sum_{q=1}^d\cos{p_{q}a}\right]
\nonumber \\
&=&
\frac {N^{2}_{f}N_s}{(2\pi)^{d}}\int_{-\pi/a}^{\pi/a}d^{d}p
\left[
\frac {1}{\sqrt {1-
\big[
\frac { {K \over a}\sum_{q=1}^d\cos{p_{q}a}}
{M+ {Kd \over a} \left( 1-r^{2} \right)- {Kr^{2} \over
a}\sum_{q=1}^d\cos{p_{q}a}}}
\big]^2
}-1
\right]
\nonumber \\
&\to^{M\to M_{c}}&
\frac {N^{2}_{f}N_s}{(2\pi)^{d}}\int_{-\pi/a}^{\pi/a}d^{d}p
\left[
\frac {1}{\sqrt {1-
\big[
\frac {\frac {K}{a}\sum_{q=1}^d\cos{p_{q}a}}
{\frac {6Kr^{2}}{a}+\frac {Kd}{a} \left( 1-r^{2} \right)
-\frac {Kr^{2}}{a}\sum_{q=1}^d\cos{p_{q}a}}}
\big]^2
}-1
\right]
\nonumber \\
&=& \frac {N^{2}_{f}N_s}{(2\pi)^{d}}\int_{-\pi/a}^{\pi/a}d^{d}p
\left[
\frac {1}{\sqrt {1-
\big[
\frac {\sum_{q=1}^d\cos{p_{q}a}}{6r^{2}+d \left(1-r^{2}\right)-
r^{2}\sum_{q=1}^d\cos{p_{q}a}}}
\big]^2
}-1\right],
\nonumber \\
\langle \bar{\psi}\psi\rangle_{V}&=&
\frac {N^{2}_{f}N_s}{(2\pi)^{d}} \sum_j \int_{-\pi/a}^{\pi/a}d^{d}p
\left[
\frac {1}{\sqrt {1-
\big[
\frac {\frac {K}{a} \left( \sum_{q=1}^d \cos{p_{q}a}-2\cos{p_{j}a} \right) }
{M+\frac {Kd}{a} \left( 1-r^{2} \right)
-\frac {Kr^{2}}{a}\sum_{q=1}^d\cos{p_{q}a}}}
\big]^2
}-1
 \right]
\nonumber \\
&\to^{M\to M_{c}}&
\frac {N^{2}_{f}N_s}{(2\pi)^{d}}
\sum_j \int_{-\pi/a}^{\pi/a}d^{d}p
\left[
\frac {1}{\sqrt {1-
\big[
\frac {\sum_{q=1}^d\cos{p_{q}a}-2\cos{p_{j}a}}{6r^{2}
+ d \left( 1-r^{2} \right) - r^{2}\sum_{q=1}^d\cos{p_q a}}}
\big]^2
}-1
\right].
\label{fortysixeth}
\end{eqnarray}
Due to the Wilson term, ${\bar \psi} \psi$ mixes with the identity operator
\cite{luoandchen}
\begin{equation}
\bar{\psi}\psi^{continuum}=\bar{\psi}\psi^{lattice}+C^{I}I.
\label{fortyseven}
\end{equation}
This relation is useful for comparing the lattice result with the continuum
theory
when $1/g^2>>1$,
but the computation of the mixing coefficient $C^I$ is quite involved.

\section{Discussions and Expectation}
\label{discussion}
In the preceding sections, we have investigated
(d+1)-dimensional Hamiltonian LGT with Wilson fermions.
Using a strong-coupling perturbative expansion,
the effective Hamiltonian in the strong-coupling regime
has been obtained and diagonalized exactly
by Bogoliubov transformation.
Some interesting physical results
for the vacuum energy, meson masses and fermion condensate
have been obtained.
For $r=0$, our results reduce to those of Luo and Chen\cite{luo}.
We will apply these techniques to lattice QCD at finite chemical potential.

\bigskip
\bigskip

\noindent
{\bf Acknowledgments}

We thank E. Gregory and S.H. Guo for useful discussions.
X.Q.L. is supported by the
National Science Fund for Distinguished Young Scholars (19825117),
Guangdong Natural Science Foundation (990212), Ministry of Education, and
Hong Kong Foundation of the Zhongshan University
Advanced Research Center.
Y.Z.F. is supported partly by the National Natural Science Foundation
(19905016).

\end{document}